\documentclass[CEJP,DVI]{cej} 
\usepackage{layout}
\usepackage{amsmath}
\usepackage{textcomp}

\title{Mixed spin-1/2 and spin-1 Ising model with uniaxial and biaxial single-ion
       anisotropy on  Bethe lattice}

\articletype{Research Article}

\author{Cesur~Ekiz\inst{1,2}\email{ekizc@yahoo.com},
        Jozef Stre\v{c}ka\inst{2}\email{jozef.strecka@upjs.sk},
        Michal Ja\v{s}\v{c}ur\inst{2}\email{michal.jascur@upjs.sk}}

\institute{
\inst{1}{Department of Physics, Faculty of Science, Adnan Menderes University, 09010 Ayd\i n, Turkey} \\
\inst{2}{Department of Theoretical Physics and Astrophysics, Faculty of Science, \\
         P. J. \v{S}af\'{a}rik University, Park Angelinum 9, 040 01 Ko\v{s}ice, Slovak Republic} \\
          }

\abstract{The mixed spin-1/2 and spin-1 Ising model on the Bethe lattice with both uniaxial as well as biaxial single-ion anisotropy terms is exactly solved by combining star-triangle and triangle-star  mapping transformations with exact recursion relations. Magnetic properties (magnetization, phase diagrams and compensation phenomenon) are investigated in detail. The particular attention is focused on the effect of uniaxial and biaxial single-ion anisotropies that basically influence the magnetic behavior of the spin-1 atoms.}

\keywords{Exact solution \*\ Uniaxial and biaxial single-ion anisotropy \*\
          Star-triangle transformation \*\ Bethe lattice}
\pacs{05.50.+q, 75.10.Hk, 75.50.Gg, 04.20.Jb}

\begin{document}
\maketitle


\section{Introduction}

In recent years, the mixed-spin Ising models belong to the most actively studied lattice-statistical models in the statistical and solid-state physics as they often exhibit unpredictable and rather
complex critical behavior. In particular, the Ising systems containing spins of different magnitudes
are still relatively simple but useful models that bring a deeper insight into the magnetic behavior
of certain ferrimagnetic materials, which are of great technological importance due to their possible applications in thermomagnetic recording \cite{man87}. With regard to this, the investigation of ferrimagnetic behavior of the mixed-spin Ising models has become a very active research field over
the last few decades. Despite the intensive studies, however, there are only few examples
of exactly solved mixed-spin Ising models, yet. Using a generalized form of the decoration-iteration
and star-triangle transformations, the mixed spin-1/2 and spin-$S$ ($S \geq 1$) Ising models on
the honeycomb, diced and decorated honeycomb lattices were exactly been treated by Fisher \cite{fis59} and Yamada \cite{yam69} many years ago. These mapping transformations were later on further
generalized in order to account also for the single-ion anisotropy terms. Actually, the procedure
based on generalized mapping transformations was recently employed to obtain exact results for
the mixed-spin Ising models with the uniaxial single-ion anisotropy on the honeycomb lattice \cite{gon85,tuc99,dak00}, bathroom-tile (4-8) lattice \cite{str06}, as well as, several decorated planar lattices \cite{jas98,dak98,str08}. On the other hand, the mixed-spin Ising models that
account also for the biaxial single-ion anisotropy have exactly been solved just on the honeycomb \cite{str04} and diced \cite{jas05} lattices, so far. It is worthy to mention that main difficulties,
which emerge when treating the Ising models refined by the biaxial single-ion anisotropy term, originate from a presence of the $x$- and $y$-components of spin operators introducing to a spin system quantum fluctuations.

Owing to these facts, the Ising models accounting for the biaxial single-ion anisotropy have recently
been a subject matter of many approximative theories as well. The ground state of the general
spin-$S$ Ising model with the biaxial single-ion anisotropy was investigated by establishing
an effective mapping to the transverse Ising model \cite{oit03} and by using the effective-field
theory with self-spin correlations \cite{jia06}. It is worthwhile to remark that the finite-temperature properties of the single-spin Ising models with the biaxial single-ion anisotropy have also been investigated within the framework of the variational mean-field treatment \cite{sou08}, the random
phase approximation \cite{tag74,mie77}, the linked-cluster series expansion \cite{pan93} and
the standard effective-field theory with correlations based on the differential operator \cite{jia00} or probability distribution \cite{edd99} techniques. Contrary to this, the effect of biaxial
single-ion anisotropy on magnetic properties of the mixed-spin Ising models was less frequently
studied and thus, it is still not fully understood, yet. Except few exactly solved cases
mentioned earlier \cite{str04,jas05}, the mixed-spin Ising models with the biaxial single-ion anisotropy term have been explored just within the conventional effective-field theory based
on the differential operator \cite{jia02} and probability distribution \cite{bel08} techniques.

With all this in mind, in this paper we will investigate the magnetic properties of the mixed-spin Ising model on the Bethe lattice with both uniaxial and biaxial single-ion anisotropy terms.
Exact solution for this model system will be obtained by combining two accurate mapping transformations
with the exact method based on recursion relations. First, the star-triangle transformation is
used to connect the mixed-spin Ising model on the Bethe lattice with the coordination number $q$=3
to its equivalent spin-1/2 Ising model on the triangular Husimi lattice. Next, the spin-1/2
Ising model on the triangular Husimi lattice is subsequently mapped by the use of triangle-star transformation to the simple spin-1/2 Ising model on the Bethe lattice. It is well known that
this latter model can be exactly treated using exact recursion relations \cite{bax82}, which will
help us to complete our exact calculation for the original mixed-spin Ising model on the Bethe lattice.

The outline of this paper is as follows. In Section 2, basic steps of the exact mapping transformations will be explained together with some details, which enable to find several exact results for the mixed-spin Ising model on the Bethe lattice with both uniaxial as well as biaxial single-ion anisotropy terms. In Section 3, the most interesting results will be presented and discussed for ground-state and finite-temperature phase diagrams, the total and sublattice magnetization. In this section, our attention will also be focused on a possibility of observing compensation phenomenon. Finally, some concluding remarks are drawn in Section 4.

\section{Model and its exact solution}

\begin{figure}
\includegraphics[width=0.8\textwidth]{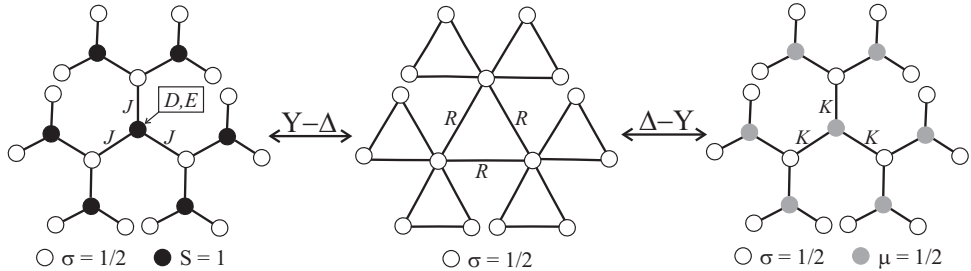}
\vspace{-0.2cm}
\caption{The figure on the left shows the mixed-spin Bethe lattice with the coordination number $q$=3, which consists of two inequivalent interpenetrating sublattices $A$ and $B$ constituted by the spin-1/2 (empty circles) and spin-1 (filled circles) atoms, respectively. The mixed-spin Ising model on the Bethe lattice is mapped via the star-triangle transformation on the spin-1/2 Ising model on the triangular Husimi lattice depicted in the central part of this figure. Subsequently, the spin-1/2 Ising model on the triangular Husimi lattice is mapped through the triangle-star transformation on the simple spin-1/2 Ising model on the Bethe lattice, which is formally divided into two equivalent interpenetrating sublattices $A$ and $B$ whose sites are shown as empty and gray circles, respectively.}
\label{fig1}
\end{figure}

Let us begin by considering a mixed spin-1/2 and spin-1 Ising model on the Bethe lattice with the coordination number $q$=3, which is schematically illustrated on the left-hand-side of Fig.~\ref{fig1}.
As shown in this figure, the mixed-spin Bethe lattice contains a central spin-1 site which has $q$ nearest-neighbor spin-1/2 sites forming the first generation. The second generation is formed by connecting $q$-1 new spin-1 sites to each site of the first generation and so on. Thus, the mixed-spin Bethe lattice consists of two inequivalent interpenetrating sublattices $A$ and $B$, which are
occupied by the spin-1/2 and spin-1 atoms depicted as empty and filled circles, respectively.
Assuming the Ising-type exchange interaction $J$ between nearest-neighboring spins from two
different sublattices $A$ and $B$, the total Hamiltonian of the investigated spin system is given by
\begin{equation}
\hat {\cal H} = -J \sum_{\langle i, j \rangle}^{2N} \hat S_i^z \hat \sigma_j^z
              - D \sum_{i = 1}^{N} (\hat S_i^z)^2
              - E \sum_{i = 1}^{N} [(\hat S_i^x)^2 - (\hat S_i^y)^2],
\label{eq1}
\end{equation}
where $N$ is a total number of lattice sites at each sublattice, $\hat S_{i}^{\alpha} (\alpha=x,y,z)$  and $\hat \sigma_{j}^{z}$ denote the spatial components of the usual spin-1 and spin-1/2 operators, respectively. The first summation is carried out over all nearest-neighboring spin pairs, while the other two summations run over all sites of the sublattice $B$. The last two terms $D$ and $E$,
which affect just the spin-1 atoms from the sublattice $B$, measure a strength of the
uniaxial and biaxial single-ion anisotropy, respectively. It should be mentioned here that by neglecting the biaxial single-ion anisotropy, i.e. setting $E$=0 in Eq. (\ref{eq1}), our model
reduces to the mixed-spin Ising model with the uniaxial single-ion anisotropy exactly solved
in several previous works \cite{sil91,alb03,eki06}. In this work, we will therefore mainly
investigate the effect of the biaxial single-ion anisotropy, which influences magnetic properties
of the model under consideration in a crucial manner. Really, the $E$ term related to the biaxial single-ion anisotropy might cause non-trivial quantum effects since it introduces the $x$ and $y$ components of spin operators into the Hamiltonian (\ref{eq1}) and thus, it is responsible for the
onset of local quantum fluctuations that are missing in the Ising models accounting for the
uniaxial single-ion anisotropy only.

Now, let us turn our attention to main points of the mapping transformation method, which enables an exact treatment of the model under investigation. By introducing site Hamiltonians, the total Hamiltonian (\ref{eq1}) can be written as a sum over all site Hamiltonians
\begin{equation}
\hat {\cal H} = \sum_{i = 1}^{N} \hat {\cal H}_{i},
\label{eq2}
\end{equation}
where each site Hamiltonian $H_{i}$ involves all the interaction terms associated with the spin-1 atom residing on the $i$th site of the sublattice $B$
\begin{equation}
\hat {\cal H}_{i} = -\hat S_{i}^{z} E_i
                  - (\hat S_{i}^{z})^2 D
                  - [(\hat S_{i}^{x})^2 - (\hat S_{i}^{y})^2] E,
\label{eq3}
\end{equation}
with $E_i  =  J (\hat \sigma_{i1}^{z} + \hat \sigma_{i2}^{z} + \hat \sigma_{i3}^{z})$. Because the Hamiltonians (\ref{eq3}) at different sites commute, i.e. $[\hat {\cal H}_i, \hat {\cal H}_j] = 0$
is valid for each $i \neq j$, the partition function of this spin system can be partially factorized and consequently rewritten in the form
\begin{equation}
{\cal Z} = \displaystyle{\sum_{\{\sigma \}}}
\prod_{i = 1}^{N} \mbox{Tr}_{S_i} \exp(- \beta \hat {\cal H}_i),
\label{eq4}
\end{equation}
where $\beta = 1/(k_B T)$, $k_B$ is Boltzmann's constant, $T$ is the absolute temperature,
$\sum_{ \{ \sigma \} }$ denotes a summation over all possible spin configurations on the sublattice A and $\mbox{Tr}_{S_i}$ stands for a trace over spin degrees of freedom of $i$th spin from the
sublattice $B$. It should be mentioned that a crucial step of our exact procedure represents
calculation of the expression  $\mbox{Tr}_{S_i} \exp(- \beta \hat {\cal H}_i)$. For this purpose,
it is useful to write the site Hamiltonian (\ref{eq3}) in the matrix form
\begin{eqnarray}
\hat {\cal H}_{i} =
\left(
\begin{array}{ccc}
-E_i-D  &  0  &   E \\
  0     &  0  &   0 \\
  E     &  0  & E_i-D  \\
\end{array}
\right),
\label{eq5}
\end{eqnarray}
in a standard basis of functions $| \pm 1 \rangle, | 0 \rangle$ corresponding, respectively,
to the three possible spin states $S_i^z = \pm 1, 0$ of $i$th atom from the sublattice $B$.
After a straightforward diagonalization of the site Hamiltonian (\ref{eq5}), the expression for
a partial trace over the spin states of $i$th spin immediately implies a possibility of performing
the generalized star-triangle mapping transformation
\begin{eqnarray}
\mbox{Tr}_{S_i} \exp(- \beta \hat {\cal H}_i) \! \! \! &=& \! \! \!
1 + 2 \exp(\beta D) \cosh \Bigl( \beta
\sqrt{J^2 (\sigma_{i1}^{z} + \sigma_{i2}^{z} + \sigma_{i3}^{z})^2 + E^2} \Bigr)
\nonumber \\
\! \! \! &=& \! \! \!
A~\exp \Bigl[ \beta R (\sigma_{i1}^{z} \sigma_{i2}^{z} + \sigma_{i2}^{z}
\sigma_{i3}^{z} + \sigma_{i3}^{z} \sigma_{i1}^{z} \bigr) \Bigr],
\label{eq6}
\end{eqnarray}
which replaces the partition function of a {\it star}, i.e. the four-spin cluster consisting
of one central spin-1 site and its three enclosing spin-1/2 sites, by the partition function
of a {\it triangle}, i.e. the three-spin cluster comprising of three spin-1/2 sites to be located
in the corners of equilateral triangle (see Fig.~\ref{fig1}). The physical meaning of the mapping (\ref{eq6}) is to remove all interaction parameters associated with the central spin-1 atom and to replace them by an effective interaction $R$ between the outer spin-1/2 atoms. It is noteworthy
that both mapping parameters $A$ and $R$ are "self-consistently" given directly by the
transformation (\ref{eq6}), which must hold for any combination of the spin states of three
spin-1/2 atoms. After taking into account all possible spin configurations of the three spin-1/2
atoms, one obtains from the star-triangle transformation (\ref{eq6}) just two independent equations
that directly determine so far not specified mapping parameters $A$ and $R$
\begin{eqnarray}
A = \Bigl(V_{1} V_{2}^3 \Bigr)^{1/4}, \qquad \qquad
\beta R = \ln \Bigl( \frac{V_1}{V_2} \Bigr).
\label{eq7}
\end{eqnarray}
In above, we have defined the functions $V_1$ and $V_2$ in order to express the transformation parameters $A$ and $R$ in a more simple and elegant form:
\begin{eqnarray}
V_{1} \! \! \! &=& \! \! \!
1 + 2 \exp( \beta D) \cosh \Bigl( \beta
\sqrt{\bigl(3J/2)^2 + E^2} \Bigr), \nonumber \\
V_{2} \! \! \! &=& \! \! \!
1 + 2 \exp( \beta D) \cosh \Bigl( \beta
\sqrt{\bigl(J/2)^2 + E^2} \Bigr).
\label{eq8}
\end{eqnarray}
When the star-triangle mapping (\ref{eq6}) is performed at each site of the sublattice $B$,
the original mixed-spin Ising model on the Bethe lattice is mapped on the spin-1/2 Ising model
on triangular Husimi lattice with the effective nearest-neighbor interaction $R$ given by the "self-consistency" condition (\ref{eq7})-(\ref{eq8}) (see Fig.~\ref{fig1}). As a matter of fact,
the substitution of the mapping transformation (\ref{eq6}) into the partition function (\ref{eq4}) establishes the relationship
\begin{equation}
{\cal Z} (\beta, J, D, E) = A^{N} {\cal Z_{\rm Husimi}}(\beta, R),
\label{eq9}
\end{equation}
between the partition function ${\cal Z}$ of the mixed-spin Ising model on the three-coordinated
Bethe lattice and respectively, the partition function ${\cal Z}_{\rm Husimi}$ of the corresponding spin-1/2 Ising model on the triangular Husimi lattice. By the triangular Husimi lattice is meant
a deep interior of the Husimi tree \cite{hus50}, which is built up from corner-sharing triangles attached to each other by their vertices as it is schematically shown in the central part of Fig.~\ref{fig1}. Even though the spin-1/2 Ising model on the triangular Husimi lattice can already
be exactly solved with the help of exact recursion relations \cite{tsu76,mon91,akh98},
it seems for us more useful to perform another mapping transformation that relates the
investigated model system to the simple spin-1/2 Ising model on the Bethe lattice for which
few exact analytical results are available in the literature \cite{bax82}.

In the next step, let us therefore perform the triangle-star transformation by inserting the spin-1/2 atom into each triangle of the triangular Husimi lattice (see Fig.~\ref{fig1}). This procedure will allow us to express the partition function of the spin-1/2 Ising model on the triangular Husimi
lattice in terms of the partition function of the simple spin-1/2 Ising model on the Bethe lattice.
However, it is more advisable to start from the Hamiltonian of the simple spin-1/2 Ising model
on the three-coordinated Bethe lattice and to show that its partition function can really be connected to the partition function of the spin-1/2 Ising model on the triangular Husimi lattice after performing the appropriate star-triangle transformation. The spin-1/2 Ising model on the Bethe lattice schematically shown on the right-hand-side of Fig.~\ref{fig1} can be defined through the Hamiltonian
\begin{equation}
{\cal H} = - K \sum_{(i,j)} \sigma_{i}^{z} \mu_{j}^{z},
\label{eq10}
\end{equation}
where we have formally distinguished the Ising spins $\sigma_{i}^{z} = \pm 1/2$ and $\mu_{j}^{z} = \pm 1/2$ to be shown as empty and gray circles, respectively, in order to divide the three-coordinated Bethe lattice into two equivalent interpenetrating sublattices $A$ and $B$.
Then, the procedure that leads to the star-triangle mapping transformation can be repeated
once again. The total Hamiltonian (\ref{eq10}) can be firstly rewritten as a sum of $N$ site Hamiltonians
\begin{equation}
{\cal H} = \sum_{j \in B}^{N} {\cal H}_{j},
\label{eq11}
\end{equation}
where each site Hamiltonian $H_{j}$
\begin{equation}
{\cal H}_{j} = - \mu_{j}^{z} K(\sigma_{j1}^{z} + \sigma_{j2}^{z} + \sigma_{j3}^{z})
\label{eq12}
\end{equation}
involves all the interaction terms of $j$th spin from the sublattice $B$. With regard to the above
definition of site Hamiltonians (\ref{eq12}), it is possible to write the following expression
for the partition function ${\cal Z}_{\rm Bethe}$ of the spin-1/2 Ising model on the Bethe lattice
\begin{equation}
{\cal Z}_{\rm Bethe} = \displaystyle{\sum_{\{\sigma\}}}
                       \prod_{j = 1}^{N} \sum_{\mu_j = \pm 1/2} \exp(- \beta \hat {\cal H}_j),
\label{eq13}
\end{equation}
where the symbol $\sum_{\{ \sigma \}}$ denotes a summation over all possible spin configurations
on the sublattice $A$ and the second summation $\sum_{\mu_j}$ is carried out over particular spin states of $j$th spin from the sublattice $B$. When this particular summation is explicitly carried
out, one gains the expression that in turn implies a possibility of performing the usual
star-triangle mapping transformation \cite{fis59}
\begin{eqnarray}
\sum_{\mu_j = \pm 1/2} \exp(- \beta \hat {\cal H}_j) =
2 \cosh \Bigl[\frac {\beta K}{2} (\sigma_{j1}^{z} + \sigma_{j2}^{z} + \sigma_{j3}^{z}) \Bigr]
= B \exp \left[ \beta R \left(\sigma_{j1}^{z} \sigma_{j2}^{z} + \sigma_{j2}^{z} \sigma_{j3}^{z}
                            + \sigma_{j3}^{z} \sigma_{j1}^{z} \right) \right]
\label{eq14}
\end{eqnarray}
with so far not specified mapping parameters $B$ and $R$. The self-consistency condition yields
for the mapping parameters $B$ and $R$ the following relations
\begin{eqnarray}
B = 2 \sqrt[4]{\cosh \left(\frac{3}{4} \beta K \right) \cosh^3 \left(\frac{1}{4} \beta K \right)},
\qquad
\beta R = \ln \left[ 2 \cosh (\frac{1}{2} \beta K) - 1 \right].
\label{eq14m}
\end{eqnarray}
When the mapping transformation (\ref{eq14}) with appropriately chosen mapping parameters (\ref{eq14m}) is performed at each site of the sublattice $B$, the spin-1/2 Ising model on the three-coordinated
Bethe lattice with the nearest-neighbor interaction $K$ is mapped onto the spin-1/2 Ising model
on the triangular Husimi lattice with the effective nearest-neighbor interaction $R$. By substituting  the transformation formula (\ref{eq14}) into the partition function (\ref{eq13}), one indeed obtains
a rather simple mapping relationship between the partition functions of the spin-1/2 Ising model on the three-coordinated Bethe lattice and respectively, the spin-1/2 Ising model on triangular Husimi lattice
\begin{equation}
{\cal Z}_{\rm Bethe} (\beta, K) = B^{N} {\cal Z_{\rm Husimi}} (\beta, R).
\label{eq15}
\end{equation}
Of course, it is also possible to find inverse triangle-star transformation that expresses the partition function of the spin-1/2 Ising model on triangular Husimi lattice in terms of that one for the corresponding spin-1/2 Ising model on the three-coordinated Bethe lattice, namely, Eq.~(\ref{eq15}) can simply be inverted. Consequently, the partition function of the mixed spin-1/2 and spin-1 Ising model
on the Bethe lattice can also be straightforwardly calculated from the partition function of the corresponding spin-1/2 Ising model on the Bethe lattice. After eliminating the partition function
of the spin-1/2 Ising model on triangular Husimi lattice from Eqs.~(\ref{eq9}) and (\ref{eq15}),
one actually obtains a simple mapping relationship
\begin{equation}
{\cal Z} (\beta, J, D, E) = \left( \frac{A}{B} \right)^{N} {\cal Z_{\rm Bethe}} (\beta, K),
\label{eq16}
\end{equation}
which relates the partition function of the mixed spin-1/2 and spin-1 Ising model on the Bethe lattice to that one of the corresponding spin-1/2 Ising model on the Bethe lattice. Besides, the effective nearest-neighbor interaction of the corresponding spin-1/2 Ising model on the three-coordinated
Bethe lattice can readily be obtained by eliminating the effective interaction $\beta R$ from Eqs.~(\ref{eq7}) and (\ref{eq14m}). This procedure enables to express the mapping parameter $K$,
which determines a strength of the nearest-neighbor interaction of the corresponding spin-1/2
Ising model on the Bethe lattice, solely as a function of the temperature and interaction
parameters $J$, $D$ and $E$ originally included in the Hamiltonian (\ref{eq1})
\begin{eqnarray}
\beta K = 2 \ln \left[\frac{V_1 + V_2}{2V_2}\pm\sqrt{\left(\frac{V_1 + V_2}{2V_2}\right)^2 - 1} \right].
\label{eq16m}
\end{eqnarray}
The sign ambiguity to emerge in the previous equation reflects the zero-field invariance of the spin-1/2 Ising model on the Bethe lattice with respect to the transformation $K \to -K$, because the relevant sign change does not affect a strength (absolute value) of the effective coupling $\beta K$. As a result, the mixed spin-1/2 and spin-1 Ising model on the Bethe lattice can alternatively be mapped either to the ferromagnetic (plus sign) or the antiferromagnetic (minus sign) spin-1/2 Ising model on the Bethe lattice, which both have the identical critical temperature $\beta_c |K| = 2 \ln 3$ for the same strength of the effective interation (from here onward, we will only utilize the solution with plus sign for simplicity). It is worthwhile to remark, moreover, that the above equation in fact completes our exact calculation, since the partition function of the
mixed spin-1/2 and spin-1 Ising model on the Bethe lattice with both uniaxial as well as biaxial single-ion anisotropy terms can be now extracted from the well-known exact result for the partition function of the corresponding spin-1/2 Ising model on the Bethe lattice \cite{bax82} unambiguously given by the effective nearest-neighbor interaction (\ref{eq16m}).

Exact results for phase diagrams and other relevant physical quantities now follow straightforwardly.
For instance, it can be easily understood from the mapping relation (\ref{eq16}) that
the mixed-spin- Ising model on the Bethe lattice becomes critical if and only if the corresponding spin-1/2 Ising model on the Bethe lattice becomes critical as well, since the mapping parameters $A$ and $B$ given by Eqs.~(\ref{eq7})-(\ref{eq8}) and (\ref{eq14m}) are analytic function in
the whole region of the interaction parameters. From this point of view, it is sufficient to compare
the effective nearest-neighbor interaction (\ref{eq16m}) of the spin-1/2 Ising  model on the Bethe
lattice with its critical value ($\beta_c K = 2 \ln 3$)
\begin{eqnarray}
\frac{V_1^c + V_2^c}{2 V_2^c} + \sqrt{\left(\frac{V_1^c + V_2^c}{2 V_2^c} \right)^2 - 1} = 3
\label{eq16c}
\end{eqnarray}
in order to determine a critical behavior of the model under investigation (the superscript $c$ in the above equation means that the inverse critical temperature $\beta_c = 1/(k_B T_c)$ enters into the  expressions $V_1^c$ and $V_2^c$ given by Eq.~(\ref{eq8}) instead of $\beta$). Similarly, the mapping relations can also be used to determine the total and sublattice magnetizations. It can be easily proved by using the mapping theorems developed by Barry \textit{et al}. \cite{bar88} that the sublattice magnetization $m_A$ of the mixed-spin Ising model on the Bethe lattice is directly
equal to the single-site magnetization of the corresponding spin-1/2 Ising model on the triangular Husimi lattice given by the effective interaction $R$, as well as, the single-site magnetization
of the corresponding spin-1/2 Ising model on the Bethe lattice given by the effective interaction $K$
\begin{equation}
m_{A} \equiv \langle \hat \sigma_{i}^z \rangle = \langle \hat \sigma_{i}^z \rangle_{{\rm Husimi}, R} =
\langle \hat \sigma_{i}^z \rangle _{{\rm Bethe}, K} \equiv m_{{\rm Bethe}}.
\label{eq17}
\end{equation}
In above, the symbols $\langle \ldots \rangle $, $\langle \ldots \rangle_{{\rm Husimi}, R}$
and $\langle \ldots \rangle _{{\rm Bethe}, K}$ denote canonical ensemble average performed
within the mixed spin-1/2 and spin-1 Ising model on the Bethe lattice, the spin-1/2 Ising model
on the Husimi lattice with the effective interaction $R$ and the spin-1/2 Ising model on the Bethe lattice with the effective interaction $K$, respectively. From this point of view, the sublattice
magnetization $m_A$ can easily be obtained from the iteration procedure based on the exact
recursion relations yielding
\begin{equation}
m_{A}  =  \frac{1}{2} \left( \frac{1 - x^3}{1 + x^3} \right),
\label{eq17a}
\end{equation}
where $x$ is given by the stable fixed point of the recurrence relations \cite{bax82}
\begin{equation}
x_{n}  =  \frac{\exp(-\frac{\beta K}{4}) + \exp(\frac{\beta K}{4}) x_{n-1}^2}
               {\exp(\frac{\beta K}{4}) + \exp(-\frac{\beta K}{4}) x_{n-1}^2}.
\label{eq17b}
\end{equation}
On the other hand, the sublattice magnetization $m_B$ can be obtained
after straightforward but a little bit cumbersome calculation from the exact Callen-Suzuki
identity \cite{cal63}
\begin{equation}
m_{B} \equiv \langle \hat S_i^{z} \rangle  =
\left \langle \frac{\mbox{Tr}_{S_i} \hat S_i^{z} \exp(- \beta \hat {\cal H}_i)}
                   {\mbox{Tr}_{S_i} \exp(- \beta \hat {\cal H}_i)} \right \rangle,
\label{eq17m}
\end{equation}
which enables to express the sublattice magnetization $m_B$ in terms of the formerly derived
sublattice magnetization $m_A$ and the triplet correlation function
$t_A \equiv \langle \hat \sigma_{i1}^z \hat \sigma_{i2}^z \hat \sigma_{i3}^z \rangle$
\begin{eqnarray}
m_B = \frac{3}{2} m_A [F(\frac{3}{2}) + F(\frac{1}{2})] + 2 t_A [F(\frac{3}{2}) - 3 F(\frac{1}{2})].
\label{eq18}
\end{eqnarray}
For completeness, the function $F(x)$ is defined as follows
\begin{eqnarray}
F(x) = \frac{Jx}{\sqrt{(Jx)^2 + E^2}}
       \frac{2\exp(\beta D) \sinh(\beta \sqrt{(Jx)^2 + E^2})}
            {1+ 2\exp(\beta D) \cosh(\beta \sqrt{(Jx)^2 + E^2})}
\label{eq19}
\end{eqnarray}
and the triplet correlation function, which is required for a computation of the sublattice magnetization $m_{B}$, can again be calculated within the framework of exact mapping theorems \cite{bar88}. The exact mapping theorems give for the unknown triplet correlation function
on the Bethe lattice the following result
$t_A \equiv \langle \hat \sigma_{i1}^z \hat \sigma_{i2}^z \hat \sigma_{i3}^z \rangle = \langle \hat \sigma_{i1}^z \hat \sigma_{i2}^z \hat \sigma_{i3}^z \rangle_{{\rm Husimi}, R} = \langle \hat \sigma_{i1}^z \hat \sigma_{i2}^z \hat \sigma_{i3}^z \rangle_{{\rm Bethe}, K} \equiv t_{\rm Bethe}$
and they also connect it to the single-site magnetization through the relation
\begin{equation}
t_{\rm Bethe}= m_{\rm Bethe} \frac{2-3 [G(\frac{3}{2})+G(\frac{1}{2})]}
                                  {4[G(\frac{3}{2}) - 3 G(\frac{1}{2})]}.
\label{eq20}
\end{equation}
In above, the newly defined function is $G(x)=\frac{1}{2} \tanh(\frac{\beta K}{2}x)$.

\section{Results and discussion}

Although all derivations presented in the preceding section hold for the ferromagnetic ($J > 0$) as well as ferrimagnetic ($J < 0$) version of the model under investigation, in what follows we shall restrict ourselves only to an analysis of the ferrimagnetic model with $J < 0$. It is worthy to mention, nevertheless, that phase diagrams displayed below for the ferrimagnetic model are valid without any changes also for the ferromagnetic model as a result of an invariance of the mapping transformations with respect to $J \to -J$ interchange.

\begin{figure}
\includegraphics[width=0.6\textwidth]{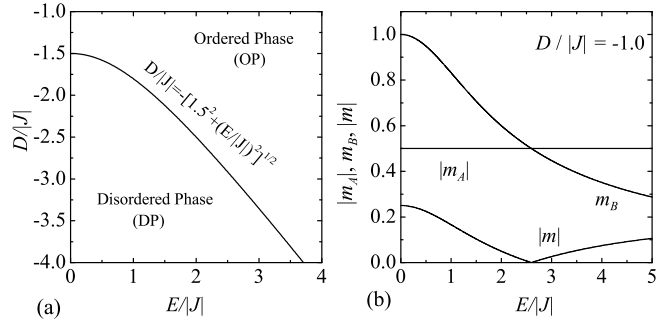}
\vspace{-1cm}
\caption{a) Ground-state phase diagram in the $E-D$ plane;
b) Single-site magnetizations versus the biaxial single-ion anisotropy $E/|J|$
and $D/|J| = -1.0$ at $T = 0$.}
\label{fig2}
\end{figure}

First, let us take a closer look at the ground-state behavior of the mixed spin-1/2 and spin-1 Ising model on the Bethe lattice. Taking into account the zero temperature limit ($T \rightarrow {0}^{+}$), one finds the following condition for a first-order phase transition line separating the magnetically ordered phase (OP) and the disordered phase (DP):
\begin{equation}
\frac{D}{|J|} = -\sqrt{\Bigl( \frac32 \Bigr)^2 + \Bigl( \frac{E}{|J|} \Bigr)^2}.
\label{eq21}
\end{equation}
The detailed analysis shows that the spin ordering appearing within OP and DP
can unambiguously be defined through the following eigenfunctions
\begin{eqnarray}
| \mbox{OP} \rangle \! \! \! &=& \! \! \! \prod_{j \in A} | - \frac{1}{2} \rangle
                      \prod_{i \in B} \left[\cos \phi \, | 1 \rangle
                                           - \sin \phi \, | -1 \rangle \right], \label{eq21e} \\
| \mbox{DP} \rangle \! \! \! &=& \! \! \! \prod_{j \in A} | \pm \frac{1}{2} \rangle
                      \prod_{i \in B} | 0 \rangle, \label{eq21f}
\end{eqnarray}
where the former products are carried out over all the spin-1/2 sites from the sublattice $A$,
the latter ones run over all the spin-1 sites from the sublattice $B$ and the angle $\phi$
determines probability amplitudes for the $|+1 \rangle$ and $| -1 \rangle$ spin states
within OP through the relation $\phi = \frac{1}{2} \arctan \left( \frac{2E}{3|J|} \right)$.
It can easily be understood from Eq.~(\ref{eq21f}) that all the spin-1 atoms reside within
DP their "non-magnetic" spin state $| 0 \rangle$ and as a result of this, all the spin-1/2
atoms become completely disordered, i.e. they are completely randomly in one of their two
possible spin states $| \pm \frac{1}{2} \rangle$. However, the more striking situation
emerges within OP, where the spin-1 atoms are in a quantum superposition of two spin states
$|+1 \rangle$ and $| -1 \rangle$ and simultaneously, all the spin-1/2 atoms reside their
"down" spin state $| - \frac{1}{2} \rangle$. Obviously, the biaxial single-ion anisotropy
enhances the probability amplitude of the spin state $| -1 \rangle$ and thus, the
$E$ term effectively lowers the sublattice magnetization $m_B$ due to the quantum reduction
of the magnetization (the sublattice magnetization $m_A$ is not directly affected by this term).
In agreement with the aforedescribed ground-state analysis, the following analytical expressions
for the single-site magnetization ($m_{A}$, $m_{B}$) and the total single-site magnetization
$m = (m_A + m_B)/2$ follow from Eqs. (\ref{eq17})-(\ref{eq20}) at the ground state:
\begin{eqnarray}
\mbox{OP:} \qquad m_A = - \frac{1}{2}, \qquad
m_B = \frac{\frac{3}{2}}{\sqrt{\Bigl( \frac{3}{2} \Bigr)^2 + \Bigl( \frac{E}{|J|} \Bigr)^2}},
\qquad m = - \frac{1}{4} + \frac{\frac{3}{4}}{\sqrt{\Bigl( \frac{3}{2} \Bigr)^2 + \Bigl( \frac{E}{|J|} \Bigr)^2}};
\label{eq22}
\end{eqnarray}
\begin{eqnarray}
\mbox{DP}: \qquad m_A = 0.0, \qquad m_B = 0.0, \qquad m = 0.0.
\label{eq23}
\end{eqnarray}
Figs.~\ref{fig2} (a) and \ref{fig2}(b) show the ground-state phase diagram in the $E$-$D$ plane
and zero-temperature variations of the magnetization with the biaxial single-ion anisotropy when $D/|J|$=-1.0. It should be mentioned that by neglecting the biaxial single-ion anisotropy, i.e. putting $E$=0 into the condition (\ref{eq21}), one recovers the phase boundary between OP and DP for the uniaxial single-ion anisotropy $D/|J|$=-1.5 that is consistent with previous calculations for this mixed-spin system on the Bethe lattice \cite{tuc99,sil91,alb03,eki06}. In this particular case,
both sublattice magnetizations have antiparallel orientation with respect to each other and
they are also both fully saturated. It should be also remarked that OP corresponds in this special
case to the simple ferrimagnetic phase. However, the ground-state behavior becomes much more complex
in the presence of the non-zero biaxial single-ion anisotropy $E$. Although the sublattice magnetization
$m_A$ remains at its saturation value in the whole OP, the sublattice magnetization $m_B$
monotonically decreases by increasing a strength of the biaxial single-ion anisotropy.
Obviously, the decrease of the sublattice magnetization $m_B$ implies a violation of
a perfect ferrimagnetic spin ordering in the OP due to the aforedescribed quantum reduction
of the magnetization caused by the $E$ term.

\begin{figure}
\includegraphics[width=0.5\textwidth]{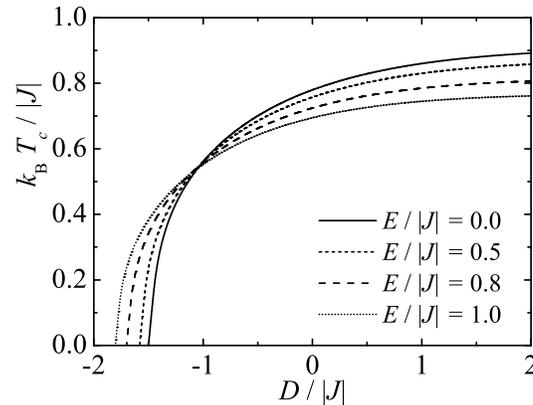}
\vspace{-1cm}
\caption{Critical temperature as a function of the uniaxial single-ion anisotropy $D/|J|$ for several values of biaxial anisotropies.}
\label{fig3}
\end{figure}

\begin{figure}
\includegraphics[width=0.5\textwidth]{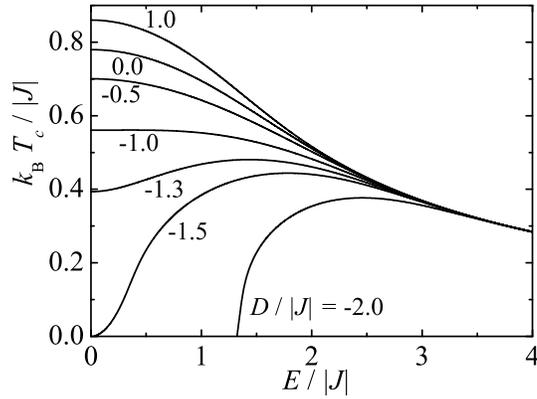}
\vspace{-1cm}
\caption{Critical temperature as a function of the biaxial single-ion anisotropy for several values of uniaxial anisotropies $D/|J|$.}
\label{fig4}
\end{figure}

Now, let us proceed to the finite-temperature properties of the system under investigation by considering the effect of uniaxial and biaxial single-ion anisotropies on the critical behavior.
The dependence of the critical temperature $T_{c}$ on the uniaxial and biaxial single-ion anisotropies is shown in Fig.~\ref{fig3} and Fig.~\ref{fig4}, respectively. In both figures, the depicted phase transition lines separate OP from DP, which always appear for temperatures above the phase transition lines. Note furthermore that the phase transitions between these two phases are of second-order with the standard mean-field like critical exponents. More specifically, Fig.~\ref{fig3} shows the critical temperature as a function of the uniaxial single-ion anisotropy for different values of the biaxial single-ion anisotropy. It can be clearly seen from this figure that the critical temperature reduces monotonically to zero by decreasing the single-ion anisotropy term $D/|J|$. For different
values of the biaxial single-ion anisotropy, the critical temperature tend to zero for the ground-state value that is consistent with the ground-state phase boundary given by Eq. (\ref{eq21}). While the uniaxial single-ion anisotropy term $D$ forces spins to lie within $x-y$ plane when $D < 0$,
the $E$ term tries to align them into the $y-z$ plane. The biaxial single-ion anisotropy thus
basically supports the magnetic ordering related to OP when $D/|J| < -1.5$ and hence, it survives
until more negative single-ion anisotropies $D/|J|$. On the other hand, the biaxial single-ion anisotropy additionally lowers the critical temperature of OP for the easy-axis single-ion anisotropy $D > 0$. It should be mentioned here that this behavior arises as a consequence of $E$ term, which induces the quantum superposition   between $| +1 \rangle$ and $| -1 \rangle$ spin states due to the nonzero quantum fluctuations
arising from this term. Altogether, it could be concluded that the quantum reduction of the magnetization also lowers the critical temperature of OP owing to the quantum fluctuations
closely associated with the $E$ term. It should be also remarked that the critical temperatures
of the mixed spin-1/2 and spin-1 Ising model with only the uniaxial single-ion anisotropy term
are recovered by neglecting the biaxial single-ion anisotropy, i.e. setting $E=0$ in Eq.~(\ref{eq1}).  As a matter of fact, our results are in this particular case consistent with those reported on previously using other exact methods for the Bethe lattice models such as the non-linear mapping \cite{sil91}, the cluster variational method in the pair approximation \cite{tuc99} or the method
based on exact recursion relations \cite{alb03,eki06}.

In order to provide a deeper insight into how the biaxial single-ion anisotropy affects the critical behavior, the critical temperature versus biaxial single-ion anisotropy dependence is shown in Fig.~\ref{fig4} for several values of the uniaxial single-ion anisotropy. Two different regions
corresponding to OP and DP are separated by second-order transition lines. Evidently, it turns
out that the biaxial single-ion anisotropy has major influence on the critical behavior of the system. The critical temperature gradually decreases with increasing  the biaxial single-ion anisotropy strength for $D/|J| > -1.0$. It is quite clear that the suppression of critical temperature can be connected to the quantum fluctuations, which occur because of the effect of $E$ term. In addition
to this rather trivial finding, one also observes the interesting nonmonotonical dependences of the critical temperature. Namely, the critical temperature firstly increases and only then gradually decreases with a strength of the biaxial single-ion anisotropy for $D/|J| < -1.0$. It should be
pointed out that the spin-1 atoms are preferably thermally excited to the $| 0 \rangle$ state
when $D < 0$, which means that they are preferably excited to the $x-y$ plane. Since the biaxial single-ion anisotropy tries to align them into the $y-z$ plane, it favors the spontaneous
long-range order along $z$ axis in that it prefers the quantum superposition between
$| \pm 1 \rangle$ spin states before the population of the $| 0 \rangle$ one. In agreement
with the aforementioned arguments, the most interesting finding to emerge here is that there
is a strong evidence for a spontaneous long-range order even under assumption of extraordinary
strong (negative) single-ion anisotropies $D/|J|$ provided that the biaxial single-ion anisotropy is strong enough. The spontaneous long-range order arises in this rather peculiar case
from the quantum fluctuations caused by the biaxial single-ion anisotropy. When comparing
the present exact results for critical temperatures with the ones for the similar mixed-spin
Ising model on the honeycomb lattice \cite{str04}, it can be concluded that the critical
temperatures for the mixed-spin Ising model on the Bethe lattice are slightly greater
than the ones for the honeycomb lattice.

\begin{figure}
\includegraphics[width=0.7\textwidth]{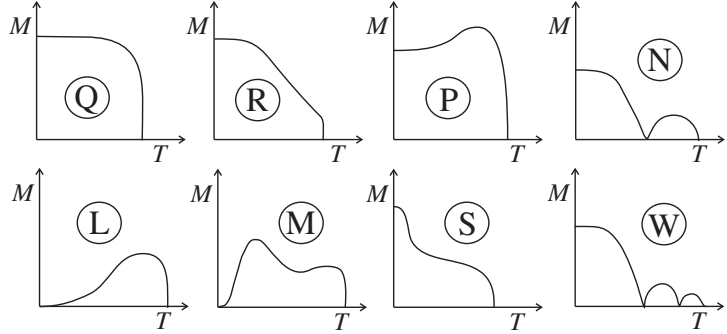}
\vspace{-10cm}
\caption{The extended N\'eel classification for different types of temperature dependences
of the total magnetization. Acronyms for particular cases are given by the circled letters.}
\label{fig5}
\end{figure}

\begin{figure}
\includegraphics[width=0.45\textwidth]{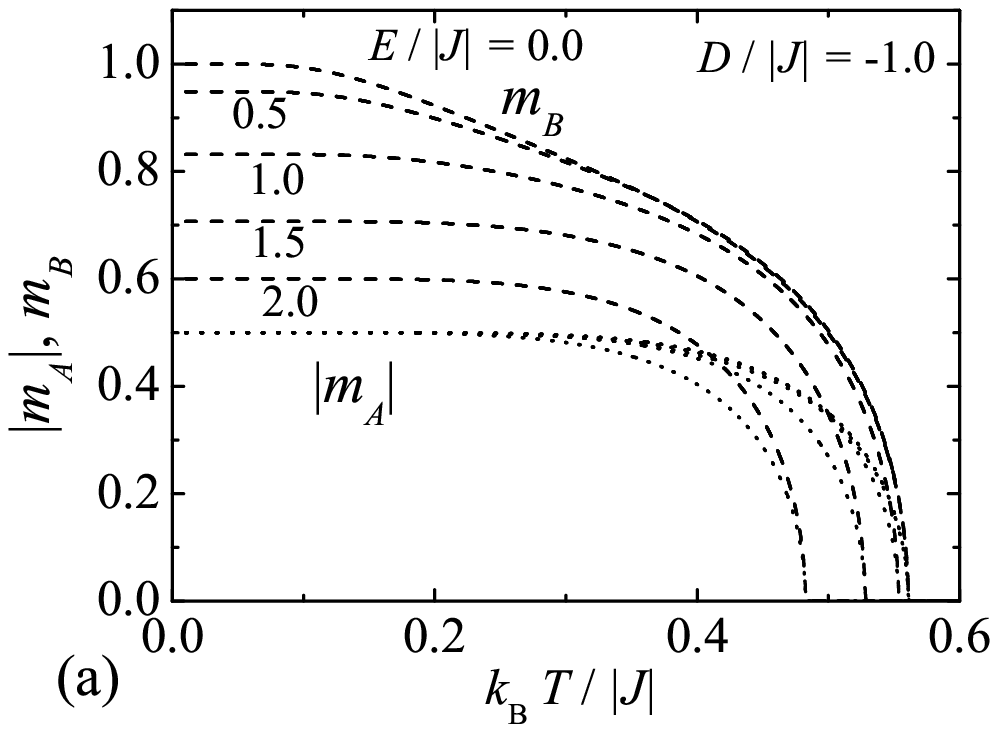}
\includegraphics[width=0.45\textwidth]{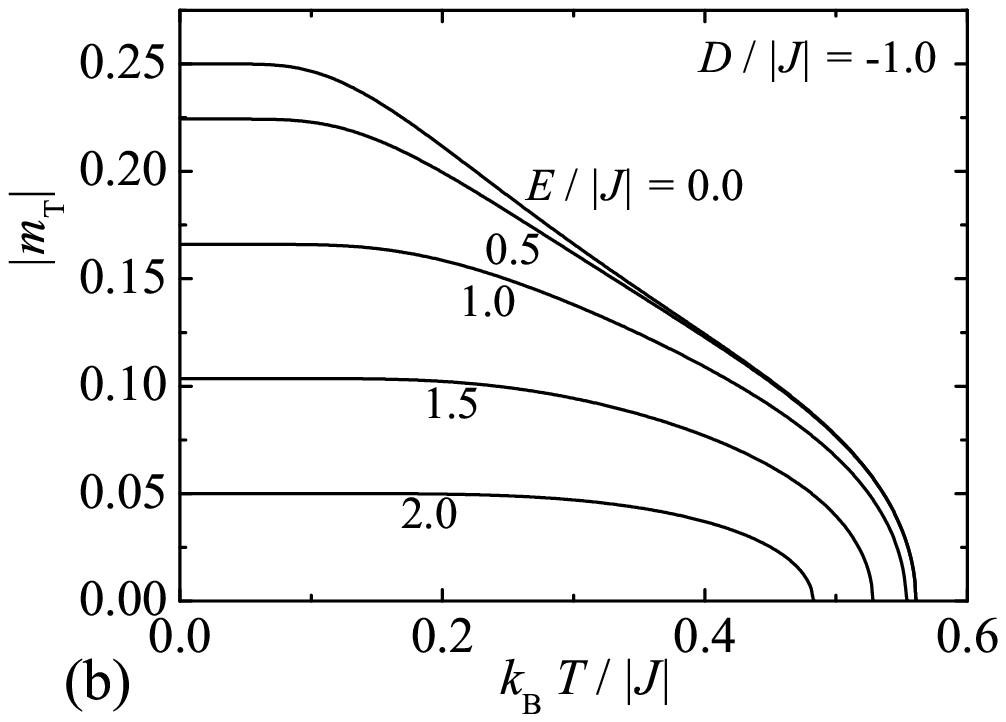}
\vspace{-0.8cm}
\caption{(a) Temperature dependences of the sublattice single-site magnetization for $D/|J| = -1.0$ and several values of the biaxial single-ion anisotropy. (b) Temperature dependences of the total magnetization normalized per one site for $D/|J| = -1.0$ and several values of the biaxial single-ion anisotropy.}
\label{fig6}
\end{figure}

Now, let us study in detail the thermal dependences of magnetization that provide an independent check of the critical behavior. For easy reference, we will further use the extended N\'eel classification
\cite{str06,nee48} to refer to different types of temperature dependences of the total magnetization, which is schematically displayed in Fig.~\ref{fig5}.
The typical thermal variations of the sublattice and total single-site magnetizations are depicted
in Fig.~\ref{fig6} for the uniaxial single-ion anisotropy $D/|J| = -1.0$ and several values of the biaxial single-ion anisotropy $E/|J|$.
As one can see from this figure, the sublattice magnetization $m_B$ varies very rapidly with $E$.
If a strength of the biaxial single-ion anisotropy is rather large, the quantum reduction of the sublattice magnetization $m_{B}$ becomes stronger due to energetical favoring of the $| - 1 \rangle$
spin state. The total magnetization then also changes from the standard R-type dependence to the more interesting Q-type dependence as it can be seen in Fig.~\ref{fig6}(b).

\begin{figure}
\includegraphics[width=0.5\textwidth]{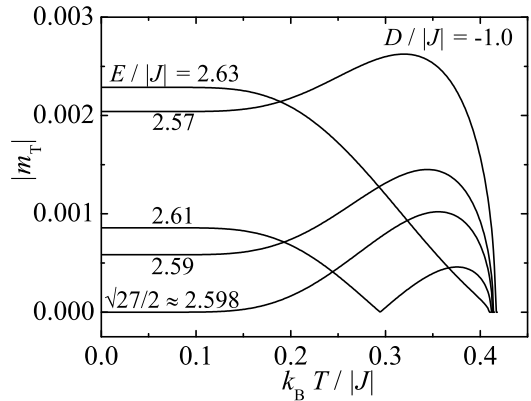}
\vspace{-1cm}
\caption{The various temperature dependences of the total magnetization normalized per one site when the strength of uniaxial single-ion anisotropy is fixed ($D/|J| = -1.0$) and the biaxial single-ion anisotropy varies in the vicinity of $E_c^0$}
\label{fig7}
\end{figure}

As far as the biaxial single-ion anisotropy at the critical value $E^0_c /|J| = \sqrt{27}/2$ is considered, the sublattice magnetization $m_{A}$ fully compensates the sublattice magnetization
$m_{B}$ in the ground-state as it can be seen from Fig.~\ref{fig2}(b) and Fig.~\ref{fig7}.
In this particular case, the sublattice magnetization $m_{B}$ becomes greater than the sublattice magnetization $m_{A}$, i.e. $|m_B|>m_A$, for all nonzero temperatures below $T_{c}$.

Finally, let us look more closely on another particular cases for different biaxial single-ion anisotropies close to the value $E^0_c$. The total magnetization exhibits the P-type curves
for $E < E_c^0$ as it is shown for $E/|J| = 2.57$ and 2.59. However, the sublattice magnetization
$|m_A|$ becomes greater than $m_{B}$ below $T_{c}$ if $E > E_c^0$ is satisfied (see $E/|J| = 2.61$
and 2.63). The more robust thermal excitations of the sublattice magnetization $m_{B}$ then
result in the N-type dependence with one compensation point if the biaxial single-ion anisotropies
are close enough but slightly higher than $E_c^0$ (for instance $E/|J| = 2.61$). It can be readily understood that the model exhibits one compensation point just in a very small restricted range of $E/|J|$ and therefore, the line of compensation temperatures is not shown for clarity in the finite-temperature phase diagrams. In the region, where the total magnetization exhibits one compensation point, the sublattice magnetization $m_A$ exceeds $m_B$ at lower temperatures,
i.e. $|m_A| > m_B$, while reverse is the case at higher temperatures, i.e. $|m_A| < m_B$
holds at higher temperatures (see $E/|J| = 2.61$). Last but not least, the R-type magnetization
curves is obtained for the more stronger biaxial single-ion anisotropy such as $E/|J| = 2.63$.

\section{Concluding remarks}

In this work, we have exactly studied the phase diagrams and magnetizations of the mixed spin-1/2 and spin-1 Ising model on the Bethe lattice ($q$=3) by the use of two precise mapping transformations and exact recursion relations. The particular attention was focused on the effect of uniaxial and biaxial single-ion anisotropies acting on the spin-1 atoms. Exact results for the phase diagrams, total and sublattice magnetizations were obtained and discussed in detail. The obtained results show that the presence of a small amount of biaxial single-ion anisotropy has a very important influence on the magnetic properties. Beside this, we have found an exact evidence that the model exhibits within OP
the remarkable quantum superposition of the $| \pm 1 \rangle$ spin states that leads to the quantum reduction of the magnetization whenever there is arbitrary but nonzero biaxial single-ion anisotropy. Macroscopically, this quantum superposition reduces the critical temperature of OP for the easy-axis uniaxial single-ion anisotropies ($D > 0$) as it also appreciably depresses the magnetization of
spin-1 atoms from its saturation value even within the ground state (see Fig.~\ref{fig2}(b)).
The discussed reduction of critical temperature as well as magnetization is obviously of purely
quantum origin, since it appears owing to the local quantum fluctuations arising from the biaxial single-ion anisotropy. On the other hand, the same quantum fluctuations can surprisingly
cause an onset of the spontaneous long-range order related to OP for the extraordinary strong easy-plane anisotropies $D<0$ provided that there is strong enough biaxial single-ion anisotropy.
Depending on a strength of the uniaxial and biaxial single-ion anisotropy parameters, the temperature variation of total magnetization has been found to be either of Q-, R-, P-, L- or N-type.
It should be also mentioned that the present results are in a good qualitative agreement with
those obtained by using the same Hamiltonian on the honeycomb lattice \cite{str04}.

Furthermore, it should be stressed that the present results are interesting both from the academic
point of view (because of their exactness) as well as from the experimental viewpoint. Namely,
the mixed-spin Ising models defined on the Bethe lattices can be useful in providing a deeper
insight into a magnetism of dendrimeric organic molecules containing radicals as their magnetic
sites \cite{mat96,cho98,ino00,rui04,aki05,he07}. In this respect, the lattice-statistical models
with spin-1/2 and spin-1 sites are the most useful ones as the most common dendrimeric organic
magnets are based on radicals such as carbenes, polyethers, trityl and trityl-anionic radicals, perchlorotriphenylmethyl radical and so on, which are basic building blocks of either
the spin-1/2 or spin-1 sites.

Finally, it is noteworthy that the exact solution of the spin-1/2 Ising model on the triangular Husimi lattice has been obtained as another by-product of our calculations. Note that this interesting spin system represents an intermediate step of our exact procedure based on the star-triangle and triangle-star mapping transformations. According to Eqs.~(\ref{eq14m}) and (\ref{eq15}), exact
results for the spin-1/2 Ising model on the triangular Husimi lattice with the nearest-neighbor interaction $R$ can simply be found from the well-known exact results for the corresponding
spin-1/2 Ising model on the Bethe lattice with the nearest-neighbor interaction $K$. For instance,
the critical temperature of the spin-1/2 Ising model on the triangular Husimi lattice can be straightforwardly obtained by putting the critical temperature $\beta_c K = 2 \ln 3$ of the
spin-1/2 Ising model on the three-coordinated Bethe lattice into the mapping relation (\ref{eq14m}), which yields after elementary calculation
\begin{eqnarray}
\beta_c R = \ln \left( \frac{7}{3} \right) \Leftrightarrow
\frac{k_B T_c}{R} = \frac{1}{\ln \left( \frac{7}{3} \right)} = 1.180225\ldots
\label{eq25}
\end{eqnarray}
As could be expected, the critical temperature of the spin-1/2 Ising model on the triangular Husimi lattice with the effective coordination number $q=6$ (i.e. the number of nearest-neighbors for each lattice site) lies in between the critical temperature of the spin-1/2 Ising model on the triangular lattice $\frac{k_B T_c}{J_t} = \frac{1}{\ln 3} = 0.910239\ldots$ and the critical temperature of the spin-1/2 Ising model on the six-coordinated Bethe lattice $\frac{k_B T_c}{K} = \frac{1}{2 \ln \left( \frac{3}{2} \right)} = 1.233151\ldots$. In this respect, the spin-1/2 Ising model on the triangular Husimi lattice slightly better approximates the spin-1/2 Ising model on the triangular lattice.
Finally, it is worthy to mention that exact procedure developed in this paper enables further remarkable extensions. For instance, the approach based on combining accurate mapping transformations
with the exact recursion relations can be further generalized to account for the general spin-$S$ problem on the sublattice $B$, the next-nearest-neighbor interaction between the spin-1/2 sites
from the sublattice $A$, the multispin exchange interaction between the spins from the sublattice
$A$ and $B$, etc. In this direction will continue our further work.

\section*{Acknowledgments}
C. Ekiz would like to thank Ministry of Education of the Slovak Republic for the award of scholarship under which part of this work was carried out. C. Ekiz gratefully acknowledges the hospitality of Department of Theoretical Physics and Astrophysics, P. J. \v{S}af\'{a}rik University.
J. Stre\v{c}ka and M. Ja\v{s}\v{c}ur acknowledge financial support provided under
the research grants LPP-0107-06 and VEGA 1/0128/08.

\end{document}